\documentclass[runningheads,a4paper]{llncs}
\usepackage{amsmath}
\usepackage{amssymb}
\usepackage{graphicx}
\usepackage{booktabs}
\usepackage{listings}
\usepackage{xcolor}
\usepackage{hyperref}
\usepackage{caption}
\usepackage{subcaption}

\usepackage{url}
\usepackage{color}
\usepackage{cite}
\usepackage{epsfig}
\usepackage{multirow}

\usepackage[nomargin,inline,marginclue,draft]{fixme}
\usepackage{cleveref}
\fxsetup{status=draft}
\fxsetup{theme=color, mode=multiuser} \FXRegisterAuthor{me}{ame}{\color{red}Me}
\newcommand{\orcid}[1]{\href{https://orcid.org/#1}{\includegraphics[scale=0.02]{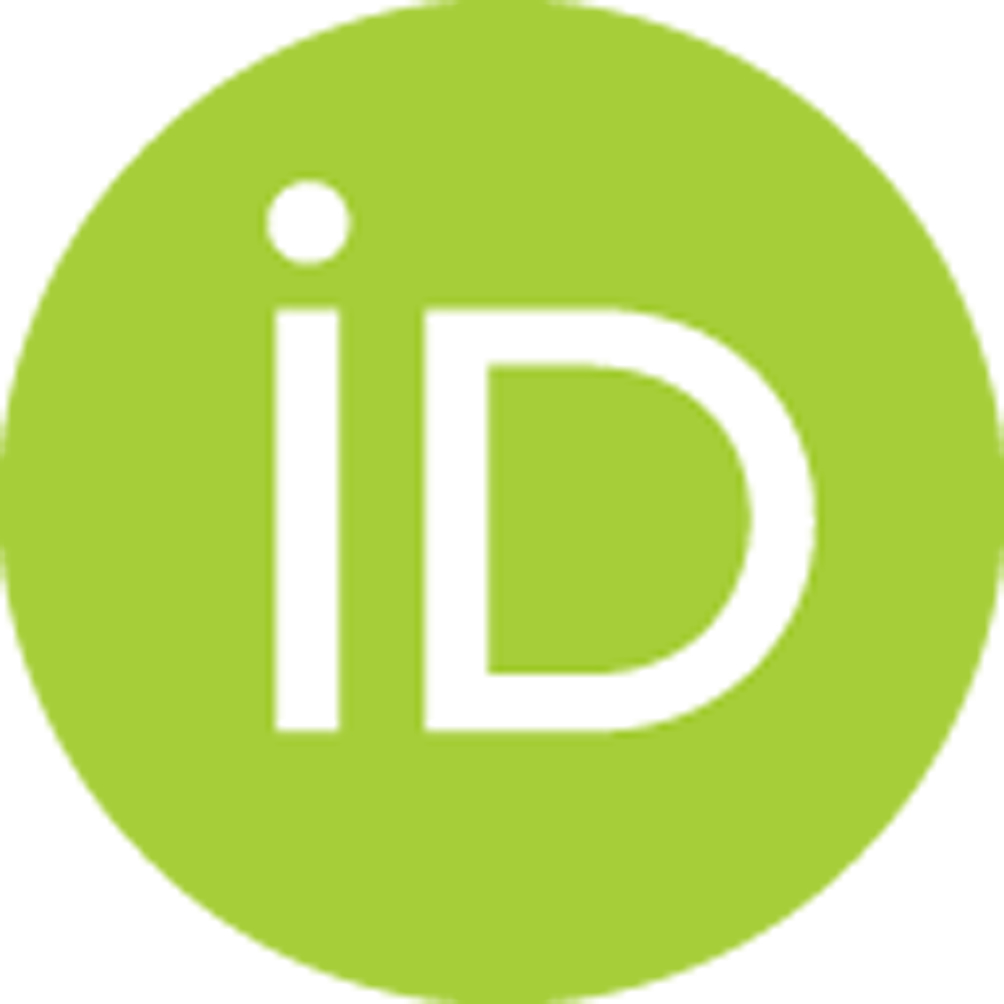}}} 

\hypersetup{
    colorlinks=true
}
\title{Federated Learning for the Classification of Tumor Infiltrating Lymphocytes}
\titlerunning{Federated Learning for the Classification of TIL}

\author{
    Ujjwal Baid\inst{1,2,3,\orcid{0000-0001-5246-2088}}
    \and
    Sarthak Pati\inst{1,2,3,4,\orcid{0000-0003-2243-8487}}
    \and
    Tahsin M. Kurc\inst{5}
    \and
    Rajarsi Gupta\inst{5,\orcid{0000-0002-1577-8718}}
    \and
    Erich Bremer\inst{5,\orcid{0000-0003-0223-1059}}
    \and
    Shahira Abousamra\inst{6\orcid{0000-0001-6214-1923}}
    \and
    Siddhesh P. Thakur\inst{1,2,3,\orcid{0000-0003-4807-2495}}
    \and
    Joel H. Saltz\inst{5,\orcid{0000-0002-3451-2165}}
    \and
    Spyridon Bakas\inst{1,2,3,*,\orcid{0000-0001-8734-6482}}
}

\authorrunning{Baid et al.}

\institute{\scriptsize{Center for Biomedical Image Computing and Analytics (CBICA), University of Pennsylvania, Philadelphia, PA, USA
    \and
    Department of Pathology and Laboratory Medicine, Perelman School of Medicine, University of Pennsylvania, Philadelphia, PA, USA
    \and
    Department of Radiology, Perelman School of Medicine, University of Pennsylvania, Philadelphia, PA, USA
    \and
    Department of Informatics, Technical University of Munich, Munich, Germany
    \and
    Department of Biomedical Informatics, Stony Brook University, Stony Brook, NY, USA
    \and
    Department of Computer Science, Stony Brook University, Stony Brook, NY, USA
}
\\
\textsuperscript{*} Corresponding author: \email{\{sbakas@upenn.edu\}}}

\begin{document}
\mainmatter
\maketitle
\setcounter{footnote}{0} 
\begin{abstract}
    We evaluate the performance of federated learning (FL)  in developing deep learning models for analysis of digitized tissue sections. A classification application was considered as the example use case, on quantifiying the distribution of tumor infiltrating lymphocytes within whole slide images (WSIs). A deep learning classification model was trained using $50\times50$ square micron patches extracted from the WSIs. We simulated a FL environment in which a dataset, generated from WSIs of cancer from numerous anatomical sites available by The Cancer Genome Atlas repository, is partitioned in 8 different nodes. Our results show that the model trained with the federated training approach achieves similar performance, both quantitatively and qualitatively, to that of a model trained with all the training data pooled at a centralized location. Our study shows that FL has tremendous potential for enabling development of more robust and accurate models for histopathology image analysis without having to collect large and diverse training data at a single location.
\end{abstract}


\section{Introduction}

    Advances in machine learning (\textbf{\textit{ML}}), and particularly deep learning (\textbf{\textit{DL}}), have shown promise in healthcare. However, the availability of large amounts of data with increased diversity is essential to produce accurate and generalizable models \cite{obermeyer2016predicting}. This is currently addressed by pooling data to a centralized location, typically facilitated via use-inspired consortia \cite{armato2004lung,thompson2014enigma,glass2018glioma,bakas2020iglass,davatzikos2020ai,habes2021brain}. However, this centralization is challenging (and at times infeasible) due to numerous concerns relating to privacy, data-ownership, intellectual property, as well as compliance with varying regulatory policies (e.g., Health Insurance Portability and Accountability Act (HIPAA) of the United States \cite{annas2003hipaa} and the General Data Protection Regulation (GDPR) of the European Union \cite{voigt2017eu}).

    In contrast to this centralized paradigm, ``federated learning'' (\textbf{\textit{FL}}) describes an approach where ML/DL models are getting trained only by sharing model updates, while all data are always retained locally within the acquiring institution \cite{mcmahan2017communication,sheller2018multiinstitutional,sheller2020federated,rieke2020future,dayan2021federated}. Several studies have demonstrated that the performance of FL models is comparable to their equivalent CL-trained models \cite{chang2018distributed,nilsson2018performance,sheller2018multiinstitutional,sheller2020federated,sarma2021federated,shen2021multi,yang2021federated}. As ML/DL methods increasingly become the primary means of analyzing large datasets, FL can offer a potential paradigm shift for multi-institutional collaborations, alleviating the need for data sharing, and hence increase access to geographically-distinct collaborators, thereby increasing the size and diversity of data used to train ML/DL models.

    Electronic capture (digitisation) and analyses of whole slide images (WSIs) of tissue specimens are becoming ubiquitous. Digital Pathology interpretation is becoming increasingly common, where many sites are actively scanning archived glass tissue slides with commercially available high speed scanners to generate high-resolution gigapixel WSIs. Alongside these efforts, a great variety of AI algorithms have been developed to extract many salient tissue and tumor characteristics from WSIs. Examples include segmentation of tumor regions, histologic subtypes of tumors, microanatomic tissue compartments; detection and classification of immune cells to identify tumor infiltrating lymphocytes (TILs); and the detection and classification of cells and nuclei. TILs are lymphoplasmacytic cells that are spatially located in tumor regions, where their role as an important biomarker for the prediction of clinical outcomes in cancer patients is becoming increasingly recognised ~\cite{mlecnik2011histopathologic,badalamenti2019role,idos2020prognostic}. Identification of the abundance and the patterns of spatial distribution of TILs in WSI represent a quantitative approach to characterizing important tumor immune interactions.

    The Stony Brook Biomedical Informatics group has actively contributed to this field of work for many years by characterizing the performance of DL pathology algorithms. Specifically, we have previously developed a DL based method to analyze WSIs to quantify distributions of TILs~\cite{saltz2018spatial,abousamra2021deep}. This method partitions a WSI into a regular mesh of image patches. Each image patch covers an area of $50\times50$ square microns, which is equivalent to $100\times100$ and $200\times200$ square pixels in a WSI captured at $20\times$ and $40\times$ magnification level, respectively. The method trains a classification model (based on a VGG16 pre-trained on ImageNet) to predict if a given image patch is TIL-positive (i.e., the patch contains 2 or more lymphocytes) or TIL-negative. This classification model has been trained and evaluated with a set of TIL-positive and TIL-negative patches from multiple cancer types with comprehensive analyses carried out to characterize performance of this method ~\cite{abousamra2021deep}.
    
    In this present study, we study the performance of federated learning using DL based detection of TILs as our model application. It is particularly useful to leverage an algorithm and WSI datasets with well known performance characteristics so that performance differences arising from FL can be easily understood. In our study, we assume that each node is responsible for all training required for slides from one or more cancer sites.  This ensures that we will see significant out-of-distribution impacts as training carried out on WSIs from a given tumor type often imperfectly generalizes to WSIs from different cancer sites.

\section{Methods}
    \subsection{Data}

    We created a training and validation dataset consisting of patches extracted from WSIs of cancer from 12 anatomical sites, comprising breast (BRCA), cervix (CESC), colon (COAD), lung (LUAD and LUSC), pancreas (PAAD), prostate (PRAD), rectum (READ), skin (SKCM), stomach (STAD), uterus (UCEC), uvea of the eye (UVM) cases, publicly available in The Cancer Genome Atlas (TCGA)~\cite{abousamra2019learning}. Tables~\ref{tab:sites}~and~\ref{tab:training_dataset} show the explicit breakdown of the dataset into individual network sites (Site 1-8). 
    
    \begin{table}[h]
    \caption{Data sharding to collaborative network sites according to anatomy.}
    \centering
    \begin{tabular}{ll}
    \hline \hline
    \textbf{Node} \hspace{10mm}       & \textbf{Anatomy}               \\ \hline \hline
    Site1                  & Cervical squamous cell carcinoma (CESC)     \\ \hline
    Site2                  & Lung squamous cell carcinoma (LUSC)         \\\hline
    \multirow{3}{*}{Site3} & Colon adenocarcinoma (COAD)                \\
                           & Pancreatic adenocarcinoma (PAAD)            \\
                           & Uterine corpus endometrial carcinoma (UCEC) \\ \hline
    Site4                  & Rectum adenocarcinoma  (READ)              \\ \hline
    Site5                  & Stomach adenocarcinoma (STAD)             \\ \hline
    Site6                  & Uveal melanoma (UVM)                      \\ \hline
    Site7                  & Lung adenocarcinoma (LUAD)                  \\ \hline
    \multirow{3}{*}{Site8} & Breast invasive carcinoma (BRCA)            \\
                           & Prostate adenocarcinoma (PRAD)              \\
                           & Skin cutaneous melanoma (SKCM)              \\ \hline \hline
    \end{tabular}
    \label{tab:sites}
    \end{table}
        
\subsection{Data Sharding}

     The complete dataset was split into vertical partitions, i.e., include all the patients of a given type of cancer and assign all the associated training data to a single site (as is shown in Table~\ref{tab:training_dataset}). With this strategy in mind, eight virtual network sites were created with each virtual site assigned to a separate computational node.

    \begin{table}[ht]
        \center
        \caption{Data split amongst different network sites for training and validation cohorts}
        \label{tab:training_dataset}
            \begin{tabular}{ccccccc}
            \hline \hline
            \textbf{Cohort} & \textbf{\begin{tabular}[c]{@{}c@{}}Total \\ Patients\end{tabular}} & \textbf{\begin{tabular}[c]{@{}c@{}}Total \\ Patches\end{tabular}} & \textbf{\begin{tabular}[c]{@{}c@{}}Patients \\ in training\end{tabular}} & \textbf{\begin{tabular}[c]{@{}c@{}}Training \\ patches\end{tabular}} & \textbf{\begin{tabular}[c]{@{}c@{}}Patients \\ in validation\end{tabular}} & \textbf{\begin{tabular}[c]{@{}c@{}}Validation \\ patches\end{tabular}} \\ \hline \hline
            Site1           & 111                                                                & 13144                                                             & 89                                                                       & 10542                                                                & 22                                                                         & 2602                                                                   \\
            Site2           & 217                                                                & 25624                                                             & 174                                                                      & 20517                                                                & 43                                                                         & 5107                                                                   \\
            Site3           & 4                                                                  & 11276                                                             & 3                                                                        & 11039                                                                & 1                                                                          & 237                                                                    \\
            Site4           & 40                                                                 & 4743                                                              & 32                                                                       & 3793                                                                 & 8                                                                          & 950                                                                    \\
            Site5           & 196                                                                & 23258                                                             & 156                                                                      & 18505                                                                & 40                                                                         & 4753                                                                   \\
            Site6           & 24                                                                 & 2438                                                              & 19                                                                       & 1938                                                                 & 5                                                                          & 500                                                                    \\
            Site7           & 10                                                                 & 23336                                                             & 8                                                                        & 16265                                                                & 2                                                                          & 7091                                                                   \\
            Site8           & 60                                                                 & 51522                                                             & 48                                                                       & 39665                                                                & 12                                                                         & 11857                                                                  \\ \hline \hline
            \end{tabular}
    \end{table}

\subsection{Model Architecture}

\begin{figure}[h]
        \centering
        \includegraphics[width=0.95\textwidth]{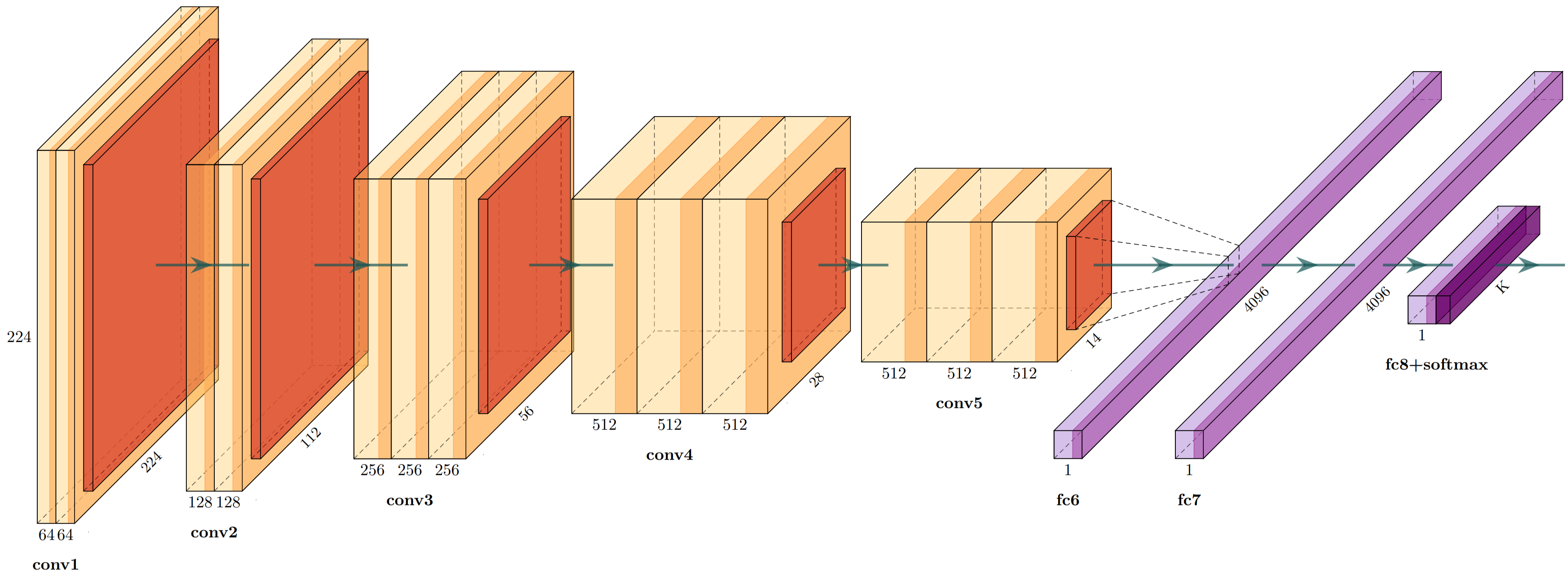}
        \caption{Deep convolutional neural network based VGG type architecture. This figure was created using PlotNeuralNetwork tool \cite{iqbal2018harisiqbal88}.}
        \label{fig:vgg}
    \end{figure}

    In this study we used the VGG network architecture \cite{simonyan2014very,ben2016fully} (Figure~\ref{fig:vgg}), which is well-known for performing classification and regression workloads, on the ImageNet classification challenge \cite{deng2009imagenet}.
    
    The VGG network used in the centralized configuration evaluated previously has 16 convolutional layers and 3 dense layers. We have modified the final classifier layers to include a global average pooling layer followed by a single dense layer, which allows greater flexibility for different types of workloads \cite{hsiao2019filter}. VGG uses $3\times3$ convolution filters and $2\times2$ max-pooling layers with a stride of 2 throughout the architecture. The original architecture uses ReLU activation function \cite{agarap2018deep} and categorical cross-entropy loss function. The initial layers of the VGG perform feature extraction and the last sigmoid layers act as the classifier.
    
    The model configuration used in the federated learning setting, was initialized using randomized weights and used as the starting point for all sites. The VGG network is trained with patch size of $300\times300$, are normalised between $[0,1]$. Sigmoid layer is used as final layer for classification. The model is trained with initial learning rate of $0.001$. The cross entropy loss with adam optimizer is used to update the weights of the model during training. 
    
\subsection{Model Training and Prediction}

    In a centralized training scenario, all of the training and validation patches are used to train a classification model. In the inference (prediction) phase, an input WSI is partitioned into patches of $50\times50$ square microns. Each patch is processed by the trained model, and the model assigns a probability value between $[0.0,1.0]$ of the patch being TIL-positive. After all of the patches are processed, a probability map is created to show the distribution of TIL patches across the WSI. 
    
    Each ``federated training round'' is defined as each site training the model on their data for a pre-defined number of epochs and providing model updates to be aggregated. For this study, the model was trained for $500$ federated rounds, where each site trained the model for a single epoch for each federated round by leveraging an independent aggregation server \cite{rieke2020future}. 

\subsection{Experimental Design}
    
    The complete study described here was facilitated by the model and training infrastructure provided by the Generally Nuanced Deep Learning Framework (GaNDLF) \cite{pati2021gandlf}. We federate \cite{sheller2018multiinstitutional,sheller2020federated,rieke2020future} the model and the entire training pipeline using the Open Federated Learning (OpenFL) library \cite{reina2021openfl}, which allows the model to be trained across multiple sites across the collaborative network without sharing any data. The quantitative performance evaluation was done using balanced classification accuracy \cite{brodersen2010balanced}.    
\section{Results}
\subsection{Quantitative Evaluation}
    
     The performance evaluation in terms of balanced classification accuracy for each individual site and the consensus model is shown in Table~\ref{results}. In the centralized training scenario we achieved accuracy of $0.75$.
    
    \begin{table}[ht]
    \centering
    \caption{Performance evaluation in terms of balanced classification accuracy for individual sites and the consensus model on the validation dataset}
    \begin{tabular}{ccccccccc|c}
    \hline \hline
    Model(row)\textbackslash{}Data(col) & Site1 & Site2 & Site3 & Site4 & Site5 & \multicolumn{1}{c}{Site6} & \multicolumn{1}{c}{Site7} & \multicolumn{1}{c}{Site8} & \multicolumn{1}{c}{Average} \\ \hline \hline
    Site 1 model & 0.90  & 0.88  & 0.78  & 0.80  & 0.86  & 0.97 & 0.86 & 0.82 & 0.86                      \\
    Site 2 model & 0.90  & 0.90  & 0.72  & 0.84  & 0.88  & 0.97 & 0.84 & 0.79 & 0.85                      \\
    Site 3 model & 0.85  & 0.84  & 0.87  & 0.82  & 0.82  & 0.98 & 0.96 & 0.84 & 0.87                      \\
    Site 4 model & 0.78  & 0.80  & 0.56  & 0.86  & 0.81  & 0.93 & 0.63 & 0.72 & 0.76                      \\
    Site 5 model & 0.88  & 0.88  & 0.70  & 0.86  & 0.88  & 0.97 & 0.80 & 0.79 & 0.85                      \\
    Site 6 model & 0.78  & 0.74  & 0.59  & 0.76  & 0.72  & 1.00 & 0.97 & 0.73 & 0.79                      \\
    Site 7 model & 0.82  & 0.79  & 0.75  & 0.76  & 0.75  & 0.97 & 0.97 & 0.80 & 0.83                      \\
    Site 8 model & 0.83  & 0.81  & 0.71  & 0.81  & 0.79  & 0.99 & 0.94 & 0.88 & 0.84                      \\
    Consensus model & 0.88  & 0.88  & 0.81  & 0.85  & 0.86  & \multicolumn{1}{c}{0.98}  & \multicolumn{1}{c}{0.95}  & \multicolumn{1}{c}{0.87} & 0.89  \\ \hline \hline
    \end{tabular}
    \label{results}
    \end{table}

\subsection{Qualitative Evaluation}

\begin{figure}[ht]
        \centering
        \includegraphics[width=0.95\textwidth]{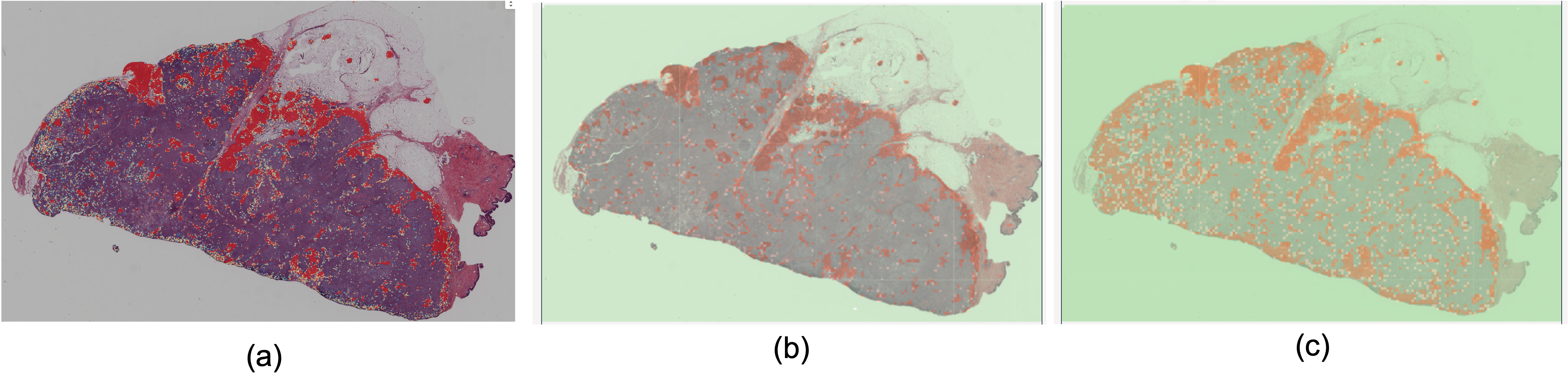}
        \caption{Predictions from different models displayed as heatmaps overlaid on the source whole slide image. (a) Predictions from the original model trained with all of the training data at a centralized location. (b) Predictions from the consensus model trained with the federated learning approach using data from all the sites. (c) Predictions from the model trained with data from a single site only.}
        \label{fig:results}
    \end{figure}

Figure~\ref{fig:results}(a) shows a visualization of patch-level probabilities from the original TIL classification model trained with all of the data in a centralized location. The patch-level probabilities are stitched together to generate a TIL heatmap overlaid on the source WSI. The RED-colored regions indicate high probability of TILs predicted by the models. The TIL predictions from the federated and site-specific models are shown in Figure~\ref{fig:results}(b) and (c), respectively. The federated consensus model predicts a TIL map that is qualitatively similar to the curated TIL map generated by the centralized model. In this example, the site-specific model predictions also appear quite similar even though this model is not trained with the entire training dataset. Even though these TIL maps look similar overall, qualitative differences are evident in terms of both the abundance and spatial distribution of TILs when comparing the federated consensus model with the site-specific model.
    
\section{Discussion}

    In this study, we generated anatomic site-specific and a federated consensus model to classify tissue regions that contain TILs. Overall, TIL maps generated with the consensus model trained in a federated manner appear comparable to the output of the original VGG16 TIL model, trained by centrally collecting all of the data in a single location \cite{saltz2018spatial}. We have also shown that there are differences in performance when the TIL models are trained with anatomic site-specific data. We evaluated performance by using held-out local validation data from each anatomic cancer site. We also performed qualitative analysis on completely unseen held-out data.
    
    The detection of TILs across WSIs of cancer was conceived based on the fairly consistent appearance of lymphocytes in normal and cancer tissues across anatomical sites. Lymphocytes are typically 8-12 um in size, round to ovoid in shape with dark blue-purple nuclei, and contain scant cytoplasm. However, detecting lymphocytes, which are called TILs within the spatial context of cancer, is challenging based on the ability of the models to distinguish TILs from cancer and other types of cells that may have overlapping features. Therefore, the original TIL model was trained with a centralized approach to provide the model with plenty of examples of non-lymphocytes and non-TILs across the complex microscopic landscapes of both normal and cancer tissues to develop a robust pan-cancer model. 
    
\begin{figure}[ht]
        \centering
        \includegraphics[width=1\textwidth]{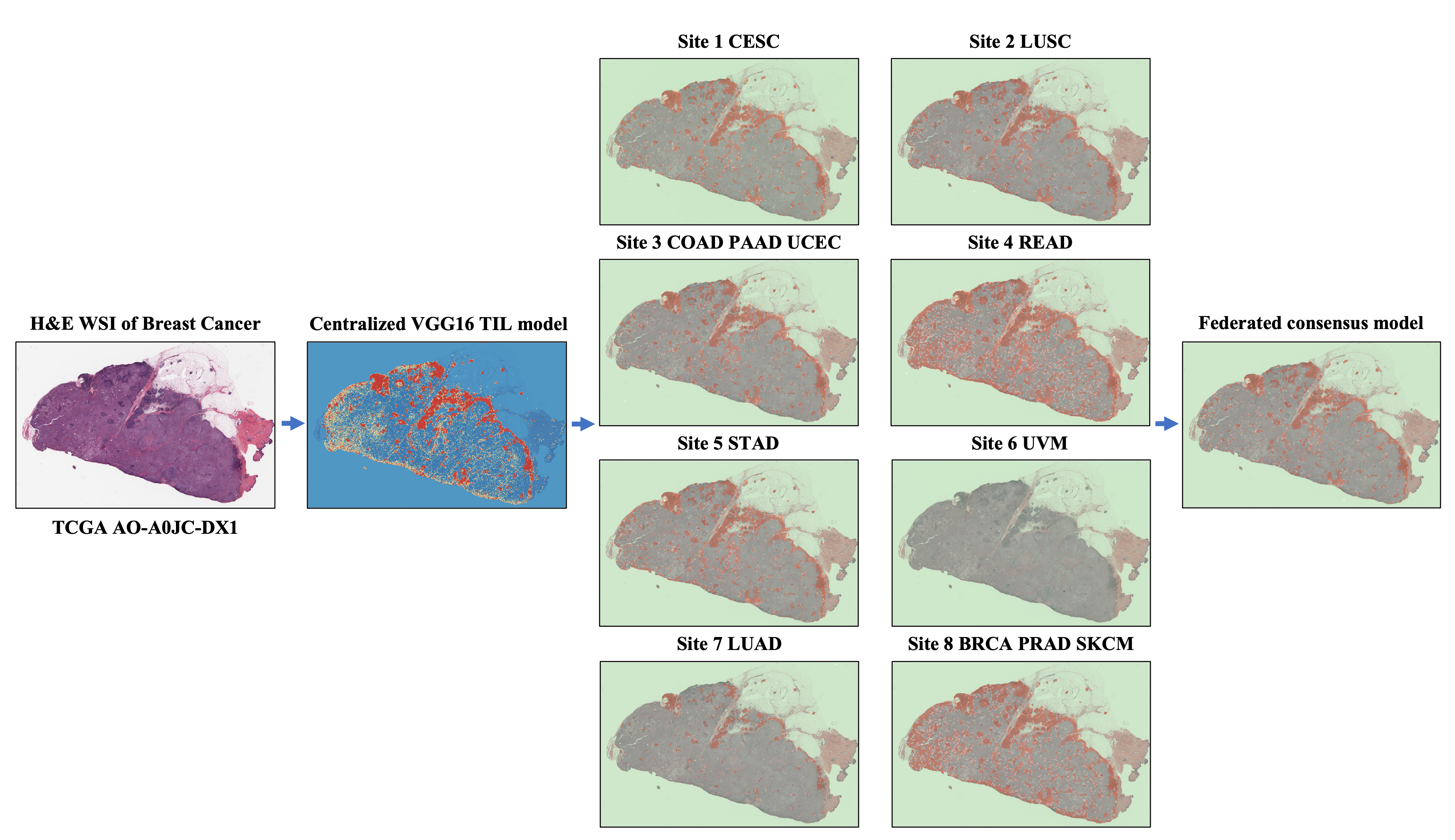}
        \caption{TIL heatmaps from federated training corresponding to the hematoxylin and eosin (H\&E) stained WSI of breast cancer, TCGA-AO-A0JC-DX1. The TIL map from the original VGG16 TIL model, trained with all of the training data at a centralized location, is shown before displaying predicted TIL maps from models trained at each anatomic site. The consensus model trained with the federated training approach using data from all the sites is shown afterwards. Visually representing the TIL maps from the various models allows qualitative evaluation of model performance for each site and consensus with respect to the original centralized VGG16 TIL model within the federated training paradigm.}
        \label{fig:results_2}
    \end{figure}

    Figure~\ref{fig:results_2} expands upon Figure~\ref{fig:results} to provide further insights into the nuances of model performance, where site-specific models (from Sites 1, 2, 3, and 5) generate TIL maps that appear quite similar. The TIL maps generated with models using training data from sites 4 and 8 also appear similar to one another with increased TILs in the left portion of the WSI at the left side of the displayed figure, which is qualitatively different from the TIL maps from site-specific models 1, 2, 3, and 5. The TIL map generated with the model from site 7 appears more conservative in comparison to models from sites 1-5 and 8, in terms of predicting less TILs globally. Notably, the model from site 6 does not predict any TILs since uveal melanoma (UVM) of the eye is typically not associated with TILs, which indicates the scarcity of TIL positive training patches. Furthermore, UVM is actually used as a negative control to test the specificity of the original TIL model in distinguishing TILs from melanoma tumor cells that can have features which mimic lymphocytes. 

    The performance of the models trained on site-specific data is dependent on the distribution of training data, specifically whether all training labels were present and if there were enough subjects in the site to warrant meaningful training. For example, it is observed (Table~\ref{results}) that Site 6 has a perfect classification performance on its own validation dataset. After further investigation we confirmed that the validation data of Site 6 includes only non-TIL patches. This shows the value of training such models in a federated setting, which allows sites that do not have enough data to be able to learn from a global consensus, thus improving their own model performance. Also, from Table 3 it can be concluded that the federated consensus model performed better than the individual sites in terms of average balanced classification accuracy. 
    
    We have evaluated the federated training paradigm using a single network architecture, and this can be expanded to include multiple architecture types to analyze which one is more suitable with regards to communication efficiency, overall convergence stability, and data privacy. Additionally, a more rigorous analysis with varying data splits is needed to provide a holistic picture of the federated training paradigm in this vertical situation. Finally, the entire study was based on simulating a federated learning paradigm on a single institution's network, but still using different computational nodes, and needs to be expanded to multiple institutions to showcase a more realistic deployment.

\section*{Funding}

 Research reported in this publication was partly supported by the National Institutes of Health (NIH) under award numbers NIH/NCI:U01CA242871, NIH/NINDS:R01NS042645, NIH/NCI:U24CA189523, NIH/NCI:UG3CA225021, NIH/NCI:4UH3CA225021, NIH/NCI:U24CA215109, and NCI/NIH:U24CA180924. The content of this publication is solely the responsibility of the authors and does not represent the official views of the NIH.

\bibliographystyle{ieeetr}
\bibliography{bibliography.bib}

\begin{thebibliography}{10}

\bibitem{obermeyer2016predicting}
Z.~Obermeyer and E.~J. Emanuel, ``Predicting the future—big data, machine
  learning, and clinical medicine,'' {\em The New England journal of medicine},
  vol.~375, no.~13, p.~1216, 2016.

\bibitem{armato2004lung}
S.~G. Armato~III, G.~McLennan, M.~F. McNitt-Gray, C.~R. Meyer, D.~Yankelevitz,
  D.~R. Aberle, C.~I. Henschke, E.~A. Hoffman, E.~A. Kazerooni, H.~MacMahon,
  {\em et~al.}, ``Lung image database consortium: developing a resource for the
  medical imaging research community,'' {\em Radiology}, vol.~232, no.~3,
  pp.~739--748, 2004.

\bibitem{thompson2014enigma}
P.~M. Thompson, J.~L. Stein, S.~E. Medland, D.~P. Hibar, A.~A. Vasquez, M.~E.
  Renteria, R.~Toro, N.~Jahanshad, G.~Schumann, B.~Franke, {\em et~al.}, ``The
  enigma consortium: large-scale collaborative analyses of neuroimaging and
  genetic data,'' {\em Brain imaging and behavior}, vol.~8, no.~2,
  pp.~153--182, 2014.

\bibitem{glass2018glioma}
T.~G. Consortium, ``{Glioma through the looking GLASS: molecular evolution of
  diffuse gliomas and the Glioma Longitudinal Analysis Consortium},'' {\em
  Neuro-Oncology}, vol.~20, pp.~873--884, 02 2018.

\bibitem{bakas2020iglass}
S.~Bakas, D.~R. Ormond, K.~D. Alfaro-Munoz, M.~Smits, L.~A.~D. Cooper,
  R.~Verhaak, and L.~M. Poisson, ``iglass: imaging integration into the glioma
  longitudinal analysis consortium,'' {\em Neuro-oncology}, vol.~22, no.~10,
  pp.~1545--1546, 2020.

\bibitem{davatzikos2020ai}
C.~Davatzikos, J.~S. Barnholtz-Sloan, S.~Bakas, R.~Colen, A.~Mahajan, C.~B.
  Quintero, J.~Capellades~Font, J.~Puig, R.~Jain, A.~E. Sloan, {\em et~al.},
  ``Ai-based prognostic imaging biomarkers for precision neuro-oncology: the
  respond consortium,'' {\em Neuro-oncology}, vol.~22, no.~6, pp.~886--888,
  2020.

\bibitem{habes2021brain}
M.~Habes, R.~Pomponio, H.~Shou, J.~Doshi, E.~Mamourian, G.~Erus, I.~Nasrallah,
  L.~J. Launer, T.~Rashid, M.~Bilgel, {\em et~al.}, ``The brain chart of aging:
  Machine-learning analytics reveals links between brain aging, white matter
  disease, amyloid burden, and cognition in the istaging consortium of 10,216
  harmonized mr scans,'' {\em Alzheimer's \& Dementia}, vol.~17, no.~1,
  pp.~89--102, 2021.

\bibitem{annas2003hipaa}
G.~J. Annas {\em et~al.}, ``Hipaa regulations-a new era of medical-record
  privacy?,'' {\em New England Journal of Medicine}, vol.~348, no.~15,
  pp.~1486--1490, 2003.

\bibitem{voigt2017eu}
P.~Voigt and A.~Von~dem Bussche, ``The eu general data protection regulation
  (gdpr),'' {\em A Practical Guide, 1st Ed., Cham: Springer International
  Publishing}, vol.~10, p.~3152676, 2017.

\bibitem{mcmahan2017communication}
B.~McMahan, E.~Moore, D.~Ramage, S.~Hampson, and B.~A. y~Arcas,
  ``Communication-efficient learning of deep networks from decentralized
  data,'' in {\em Artificial intelligence and statistics}, pp.~1273--1282,
  PMLR, 2017.

\bibitem{sheller2018multiinstitutional}
M.~J. Sheller, G.~A. Reina, B.~Edwards, J.~Martin, and S.~Bakas,
  ``Multi-institutional deep learning modeling without sharing patient data: A
  feasibility study on brain tumor segmentation,'' in {\em International MICCAI
  Brainlesion Workshop}, pp.~92--104, Springer, 2018.

\bibitem{sheller2020federated}
M.~J. Sheller, B.~Edwards, G.~A. Reina, J.~Martin, S.~Pati, A.~Kotrotsou,
  M.~Milchenko, W.~Xu, D.~Marcus, R.~R. Colen, {\em et~al.}, ``Federated
  learning in medicine: facilitating multi-institutional collaborations without
  sharing patient data,'' {\em Scientific reports}, vol.~10, no.~1, pp.~1--12,
  2020.

\bibitem{rieke2020future}
N.~Rieke, J.~Hancox, W.~Li, F.~Milletari, H.~R. Roth, S.~Albarqouni, S.~Bakas,
  M.~N. Galtier, B.~A. Landman, K.~Maier-Hein, {\em et~al.}, ``The future of
  digital health with federated learning,'' {\em NPJ digital medicine}, vol.~3,
  no.~1, pp.~1--7, 2020.

\bibitem{dayan2021federated}
I.~Dayan, H.~R. Roth, A.~Zhong, A.~Harouni, A.~Gentili, A.~Z. Abidin, A.~Liu,
  A.~B. Costa, B.~J. Wood, C.-S. Tsai, {\em et~al.}, ``Federated learning for
  predicting clinical outcomes in patients with covid-19,'' {\em Nature
  medicine}, vol.~27, no.~10, pp.~1735--1743, 2021.

\bibitem{chang2018distributed}
K.~Chang, N.~Balachandar, C.~Lam, D.~Yi, J.~Brown, A.~Beers, B.~Rosen, D.~L.
  Rubin, and J.~Kalpathy-Cramer, ``Distributed deep learning networks among
  institutions for medical imaging,'' {\em Journal of the American Medical
  Informatics Association}, vol.~25, no.~8, pp.~945--954, 2018.

\bibitem{nilsson2018performance}
A.~Nilsson, S.~Smith, G.~Ulm, E.~Gustavsson, and M.~Jirstrand, ``A performance
  evaluation of federated learning algorithms,'' in {\em Proceedings of the
  Second Workshop on Distributed Infrastructures for Deep Learning}, pp.~1--8,
  2018.

\bibitem{sarma2021federated}
K.~V. Sarma, S.~Harmon, T.~Sanford, H.~R. Roth, Z.~Xu, J.~Tetreault, D.~Xu,
  M.~G. Flores, A.~G. Raman, R.~Kulkarni, {\em et~al.}, ``Federated learning
  improves site performance in multicenter deep learning without data
  sharing,'' {\em Journal of the American Medical Informatics Association},
  vol.~28, no.~6, pp.~1259--1264, 2021.

\bibitem{shen2021multi}
C.~Shen, P.~Wang, H.~R. Roth, D.~Yang, D.~Xu, M.~Oda, W.~Wang, C.-S. Fuh, P.-T.
  Chen, K.-L. Liu, {\em et~al.}, ``Multi-task federated learning for
  heterogeneous pancreas segmentation,'' in {\em Clinical Image-Based
  Procedures, Distributed and Collaborative Learning, Artificial Intelligence
  for Combating COVID-19 and Secure and Privacy-Preserving Machine Learning},
  pp.~101--110, Springer, 2021.

\bibitem{yang2021federated}
D.~Yang, Z.~Xu, W.~Li, A.~Myronenko, H.~R. Roth, S.~Harmon, S.~Xu, B.~Turkbey,
  E.~Turkbey, X.~Wang, {\em et~al.}, ``Federated semi-supervised learning for
  covid region segmentation in chest ct using multi-national data from china,
  italy, japan,'' {\em Medical image analysis}, vol.~70, p.~101992, 2021.

\bibitem{mlecnik2011histopathologic}
B.~Mlecnik, M.~Tosolini, A.~Kirilovsky, A.~Berger, G.~Bindea, T.~Meatchi,
  P.~Bruneval, Z.~Trajanoski, W.-H. Fridman, F.~Pag{\`e}s, {\em et~al.},
  ``Histopathologic-based prognostic factors of colorectal cancers are
  associated with the state of the local immune reaction,'' {\em Journal of
  clinical oncology}, vol.~29, no.~6, pp.~610--618, 2011.

\bibitem{badalamenti2019role}
G.~Badalamenti, D.~Fanale, L.~Incorvaia, N.~Barraco, A.~Listi, R.~Maragliano,
  B.~Vincenzi, V.~Calo, J.~L. Iovanna, V.~Bazan, {\em et~al.}, ``Role of
  tumor-infiltrating lymphocytes in patients with solid tumors: Can a drop dig
  a stone?,'' {\em Cellular immunology}, vol.~343, p.~103753, 2019.

\bibitem{idos2020prognostic}
G.~E. Idos, J.~Kwok, N.~Bonthala, L.~Kysh, S.~B. Gruber, and C.~Qu, ``The
  prognostic implications of tumor infiltrating lymphocytes in colorectal
  cancer: a systematic review and meta-analysis,'' {\em Scientific reports},
  vol.~10, no.~1, pp.~1--14, 2020.

\bibitem{saltz2018spatial}
J.~Saltz, R.~Gupta, L.~Hou, T.~Kurc, P.~Singh, V.~Nguyen, D.~Samaras, K.~R.
  Shroyer, T.~Zhao, R.~Batiste, {\em et~al.}, ``Spatial organization and
  molecular correlation of tumor-infiltrating lymphocytes using deep learning
  on pathology images,'' {\em Cell reports}, vol.~23, no.~1, pp.~181--193,
  2018.

\bibitem{abousamra2021deep}
S.~Abousamra, M.~Gupta, L.~Hou, R.~Batiste, T.~Zhao, A.~Shankar, A.~Rao,
  C.~Chen, D.~Samaras, T.~Kurc, {\em et~al.}, ``Deep learning-based mapping of
  tumor infiltrating lymphocytes in whole slide images of 23 types of cancer,''
  {\em Frontiers in oncology}, p.~5971, 2021.

\bibitem{abousamra2019learning}
S.~Abousamra, L.~Hou, R.~Gupta, C.~Chen, D.~Samaras, T.~Kurc, R.~Batiste,
  T.~Zhao, S.~Kenneth, and J.~Saltz, ``Learning from thresholds: fully
  automated classification of tumor infiltrating lymphocytes for multiple
  cancer types,'' {\em arXiv preprint arXiv:1907.03960}, 2019.

\bibitem{simonyan2014very}
K.~Simonyan and A.~Zisserman, ``Very deep convolutional networks for
  large-scale image recognition,'' {\em arXiv preprint arXiv:1409.1556}, 2014.

\bibitem{ben2016fully}
A.~Ben-Cohen, I.~Diamant, E.~Klang, M.~Amitai, and H.~Greenspan, ``Fully
  convolutional network for liver segmentation and lesions detection,'' in {\em
  Deep learning and data labeling for medical applications}, pp.~77--85,
  Springer, 2016.

\bibitem{deng2009imagenet}
J.~Deng, W.~Dong, R.~Socher, L.-J. Li, K.~Li, and L.~Fei-Fei, ``Imagenet: A
  large-scale hierarchical image database,'' in {\em 2009 IEEE conference on
  computer vision and pattern recognition}, pp.~248--255, Ieee, 2009.

\bibitem{iqbal2018harisiqbal88}
H.~Iqbal, ``Harisiqbal88/plotneuralnet v1.0.0,'' Dec. 2018.

\bibitem{hsiao2019filter}
T.-Y. Hsiao, Y.-C. Chang, H.-H. Chou, and C.-T. Chiu, ``Filter-based
  deep-compression with global average pooling for convolutional networks,''
  {\em Journal of Systems Architecture}, vol.~95, pp.~9--18, 2019.

\bibitem{agarap2018deep}
A.~F. Agarap, ``Deep learning using rectified linear units (relu),'' {\em arXiv
  preprint arXiv:1803.08375}, 2018.

\bibitem{pati2021gandlf}
S.~Pati, S.~P. Thakur, M.~Bhalerao, U.~Baid, C.~Grenko, B.~Edwards, M.~Sheller,
  J.~Agraz, B.~Baheti, V.~Bashyam, {\em et~al.}, ``Gandlf: A generally nuanced
  deep learning framework for scalable end-to-end clinical workflows in medical
  imaging,'' {\em arXiv preprint arXiv:2103.01006}, 2021.

\bibitem{reina2021openfl}
G.~A. Reina, A.~Gruzdev, P.~Foley, O.~Perepelkina, M.~Sharma, I.~Davidyuk,
  I.~Trushkin, M.~Radionov, A.~Mokrov, D.~Agapov, {\em et~al.}, ``Openfl: An
  open-source framework for federated learning,'' {\em arXiv preprint
  arXiv:2105.06413}, 2021.

\bibitem{brodersen2010balanced}
K.~H. Brodersen, C.~S. Ong, K.~E. Stephan, and J.~M. Buhmann, ``The balanced
  accuracy and its posterior distribution,'' in {\em 2010 20th international
  conference on pattern recognition}, pp.~3121--3124, IEEE, 2010.

\end{thebibliography}

\end{document}